\documentclass[11pt,twoside,letterpaper]{article} 
\usepackage{times,fancyhdr,graphicx}
\def\2F1{~_2F_1}
\def\aj{AJ  }
 
\def\mnras{MNRAS\,  }

\def\za{Z. Astrophys.  } 

  

\makeatletter
\def\cleardoublepage{\clearpage\if@twoside \ifodd\c@page\else%
    \hbox{}%
    \thispagestyle{empty}%
    \newpage%
    \if@twocolumn\hbox{}\newpage\fi\fi\fi}
\makeatother

\renewcommand{\thesection}{\arabic{section}.}
\renewcommand{\thesubsection}{\thesection\arabic{subsection}.}

\def\figurename{Figure}
\makeatletter
\renewcommand{\fnum@figure}[1]{\figurename~\thefigure.}
\makeatother

\def\tablename{Table}
\makeatletter
\renewcommand{\fnum@table}[1]{\tablename~\thetable.}
\makeatother

\begin{document}
\title{
{\begin{flushleft} \vskip 0.45in
{\normalsize\bfseries\textit{Chapter~1}}
\end{flushleft}
\vskip 0.45in \bfseries\scshape  
Asteroids dimensions and the 
             Truncated  Pareto distribution 
}}
\author{\bfseries\itshape  
 Lorenzo Zaninetti \thanks{Email address: zaninetti@ph.unito.it}\\
Dipartimento di Fisica
           Generale, Via Pietro Giuria 1, 10125 Torino\\
  \bfseries\itshape  
  Mario Ferraro \thanks{Email address: ferraro@ph.unito.it}\\
  Dipartimento di Fisica
  Generale, Via Pietro Giuria 1, 10125 Torino.
}

\date{}
\maketitle \thispagestyle{empty} \setcounter{page}{1}
\thispagestyle{fancy} \fancyhead{}
\fancyhead[L]{In: Book Title \\
Editor: Editor Name, pp. {\thepage-\pageref{lastpage-01}}} 
\fancyhead[R]{ISBN 0000000000  \\
\copyright~2007 Nova Science Publishers, Inc.} \fancyfoot{}
\renewcommand{\headrulewidth}{0pt}

\vspace{2in}

\vspace{.08in} \noindent \textbf{Keywords:} 
Asteroids
\pagestyle{fancy} \fancyhead{} \fancyhead[EC]{Zaninetti and Ferraro}
\fancyhead[EL,OR]{\thepage} \fancyhead[OC]
{
 Pareto distribution   and  asteroids size
} \fancyfoot{}
\renewcommand\headrulewidth{0.5pt}

\section*{Abstract}

In this chapter first the statistics  of the standard   
and truncated Pareto distributions 
are  derived and  
used to fit empirical values of asteroids diameters  
from different  
families, namely, Koronis , 
Eos and Themis, and from the 
Astorb database.
A theoretical analysis is then carried out and  
two possible physical mechanisms 
 are suggested that account for  
Pareto tails in
distributions of asteroids  diameter.

\section{Introduction}

The interest in the study  of asteroids in the inner solar system
lays  in their connection with the formation of planets 
 and  
their temporal evolution.

Among others,  studies on asteroids formations and evolutions involve  
\begin{enumerate}

\item the effects of asteroid collisional history 
on sizes and spins of
present-day objects  \cite{Davis1989},

\item 
realistic collisional scaling
laws and the implications of including
observables, such as collisional produced families, in constraining the
collisional history of main-belt asteroids 
\cite{Davis2002b},

\item 
the creation of a model of the main asteroid 
belt whose purpose
is to describe distribution of size 
size frequency of asteroids  and simulate their  
number
\cite{Tedesco2005}.
\item  temporal evolution 
for 4.2 Myr
of
test particles, 
which were
initially placed on a perfectly rectangular grid 
and subjected  to   gravitational interactions 
with the Sun and five planets,
from Mars to Neptune , see \cite{Michtchenko2010}.
\end{enumerate}

On the other hand, it has been shown that experimental observations can be  fitted   
with a differential size distribution
\begin{equation} 
n=dN/dD=n(D) \propto D^{-\alpha} 
\quad ,
\end{equation}
where $D$  is the diameter  in Km ,
$\alpha$ the exponent  of the inverse power law 
and  $n$ the number
of asteroids comprised between   $D$ and $D+dD$.
Measurements of the properties of 13,000 asteroids 
detected by the Sloan Digital Sky Survey (SDSS)
present  a differential size distribution
that for  $D\ge 5Km $ is  $n\propto D^{-4}$  
and  for  $D\le 5Km $ is  $n\propto D^{-2.3}$,
see  \cite{ Michtchenko2010}.

The ongoing  simulations as well the observations
require a careful  analysis  of the 
Pareto distribution  \cite{Pareto,evans}
and the truncated  
Pareto  distribution   \cite{Cohen1988,Devoto1998,Aban2006}.
This paper  presents in Section~\ref{preliminar}
a comparison between the Pareto 
and the truncated Pareto distributions.
In Section~\ref{secapplications}  the theoretical results are 
 applied to the 
distribution of  the 
radius of asteroids.
Two  physical mechanisms  that produces a Pareto type
distribution for diameters   are  presented 
in Section~\ref{tails}

\section{Statistical Distribution}
\label{preliminar}
Let $X$ be a random variable taking values $x$ in the interval 
$[a, \infty]$, $a>0$.
The  probability density function (in the following pdf)
named Pareto  
is defined by  \cite{evans}
\begin {equation}
f(x;a,c) = {c a^c}{x^{-(c+1)}} \quad ,
\label{pareto}
\end {equation}
$ c~>0$,
and the Pareto distribution functions   is
$F(x:a,c)=1-a^cx^{-c}$

An upper truncated Pareto random variable is defined in the interval 
$[a,b]$ and the corresponding pdf is 
\begin {equation}
f_T(x;a,b,c ) = \frac{ca^cx^{-(c+1)}}{1-\left (\frac{a}{b}\right)^c}
\quad ,  \label{eq:pdf}
\end {equation}
\cite{Aban2006} and 
the truncated Pareto distribution function is  

\begin{equation}
F_T(x;a,b,c) =
\frac {1 -(\frac {a}{x})^c} {1-(\frac{a}{b})^c}
\quad .
\end{equation}

Momenta of the truncated distributions exist for all $c>0$. For instance,
the 
mean of   $f_T(x; a,b,c)$ is, for  $c\neq 1$ and $c=1$, respectively,
\begin{equation}
\label{eq:avtr}
<x> =
\frac {ca}{c-1} 
\frac {1 -(\frac{a}{b})^{c-1}}{1-(\frac{a}{b})^c}, \quad 
<x> =\frac{ca^c}{1-\left (\frac{a}{b}\right )^c}\ln\frac{b}{a}
\end{equation}

Similarly, if $c\neq 2$, the variance is given by
\begin{equation}
\label{eq:vart}
\sigma^2 = \frac{ca^2}{(c-2)}\frac{1-\left (\frac{a}{b} \right)^{c-2}}
{1-\left (\frac{a}{b} \right )^c} -<x>^2,
\end {equation}

whereas for $c=2$
\begin{equation}
\label{eq:svart}
\frac{ca^c}{1-\left (\frac{a}{b}\right )^c}\ln\frac{b}{a}-<x>^2.
\end{equation} 

In general the  $n-th$ central moment is  
\begin{eqnarray}
\int_a^b(x-<x>)^n f_T(x)dx=  \nonumber \\
 \left( -{\it <x>} \right) ^{n}{a}^{-c}
{\2F1(-c,-n;\,1-c;\,{\frac {a}{{\it <x>}}})} \left(  \left( {a}^{c}
 \right) ^{-1}- \left( {b}^{c} \right) ^{-1} \right) ^{-1}
\nonumber \\
- \left( -{\it <x>} \right) ^{n}{b}^{-c}
{\2F1(-c,-n;\,1-c;\,{\frac {b}{{\it <x>}}})} \left(  \left( {a}^{c}
 \right) ^{-1}- \left( {b}^{c} \right) ^{-1} \right) ^{-1}
\end{eqnarray}
where ${\2F1(a,b;\,c;\,z)}$ is a regularized hypergeometric function,
see~\cite{Abramowitz1965,Seggern1992,Thompson1997}.
An analogous formula
 based
on some of the properties of the incomplete beta function, 
see~\cite{Grad2000}  and \cite{Prud1986} ,
can be  found in~\cite{Ali2006}.
The  median  $m$  of the  Pareto  distribution is 
\begin {equation}
m=
{2}^{{c}^{-1}}a
\quad ,
\end{equation}
and the  median  of  truncated Pareto  $m_T$ 
\begin {equation}
m_T=   
a{2}^{{c}^{-1}} \left( {a}^{c}{b}^{-c}+1 \right) ^{-{c}^{-1}}
\quad .
\end{equation}

Parameters of the truncated Pareto pdf 
from empirical data can be obtained via  
the maximum likelihood method; explicit formulas 
for maximum likelihood estimators  (MLE) are given in \cite{Cohen1988}, 
and for  the more general case  in 
 \cite{Aban2006}, whose results we report here for completeness. 

Consider a random sample  ${\mathcal X}=x_1, x_2 , \dots , x_n$ and let 
$x_{(1)} \geq x_{(2)} \geq \dots \geq x_{(n)}$ denote 
their order statistics so that 
$x_{(1)}=\max(x_1, x_2, \dots, x_n)$, $x_{(n)}=\min(x_1, x_2, \dots, x_n)$.

The MLE of the parameters $a$ and $b$ 
are
\begin{equation}
{\tilde a}=x_{(n)}, \qquad {\tilde b}=x_{(1)},
\label{eq:firstpar}
\end{equation}  

respectively, and $\tilde c$ is the solution of the equation

\begin {equation}
\label{equationmle}
\frac {n}{{\tilde c}} +
\frac {n  (\frac {x_{(n)}}{x_{(1)}})^{\tilde c} 
\ln (\frac{x_{(n)}}{x_{(1)}})}{ 1-(\frac{x_{(n)}}{x_{(1)}})^{\tilde c }}
- \sum_{i=1}^n  [\ln x_i-\ln x_{(n)}]
= 0,
\end {equation}
\cite{Aban2006}.

There exists a 
simple test to see whether a Pareto model is appropriate \cite{Aban2006}:
the null hypothesis $H_0:\nu=\infty$ is rejected if and only if 
$x_{(1)} < [nC/(-\ln q)]^{1/c}$, $0<q<1$, where $C=a^c$.
The approximate $p$-value of this test is given 
by $p=\exp\left \{-nCx^{-c}_{(1)}\right \}$, and a small value of $p$ 
indicates that the 
Pareto model is not a good fit.

Given a set of data is  often difficult to decide if 
they agree more closely with $f$ or $f_T$, since, in the interval 
$[a,b]$, they differ only or a multiplicative factor 
$1-(a/b)^c$, that  approaches $1$ even for relatively small 
values of $c$ if the interval $[a, b]$
is not too small. 
For this reason,  rather than $f$ and $f_T$,  
the distributions $P(X>x)$ and $P_T(X>x)$ are used,
often called survival functions,
that are given respectively by 
\begin{equation}
\label{eq:surv}
P(X>x)=S(x)=1-F(x;a,c)=a^cx^{-c}
\end{equation}
and 
\begin{equation}
\label{eq:surt}
P_T(X>x)=S_T(x)=1-F_T(x;a,b,c)=
\frac{ca^c \left(x^{-c}-b^{-c}\right)}{1-\left(\frac{a}{b}\right)^c}.
\end{equation}
The Pareto variate $X$  can be generated by  
\begin{equation}
X  : a,c   \sim
a\left( 1 - R  \right) ^{-\frac{1}{c}}
\quad ,
\label{randompareto}
\end{equation}
and the truncated Pareto variate $X_T$ by  
\begin{equation}
X_T  : a,b,c   \sim
a\left( 1 - R ({1-(\frac{a}{b})^c)}  \right) ^{-\frac{1}{c}}
\quad ,
\label{randomparetotrunc}
\end{equation}
where $ R$ is the unit rectangular variate.

\section{Application to the asteroids}
\label{secapplications} 

We have tested the hypothesis  that diameters of asteroids follows a Pareto distribution by by considering  different families of asteroids, namely, Koronis , Eos and Themis.

\begin{figure}
{
\includegraphics[width=10cm]{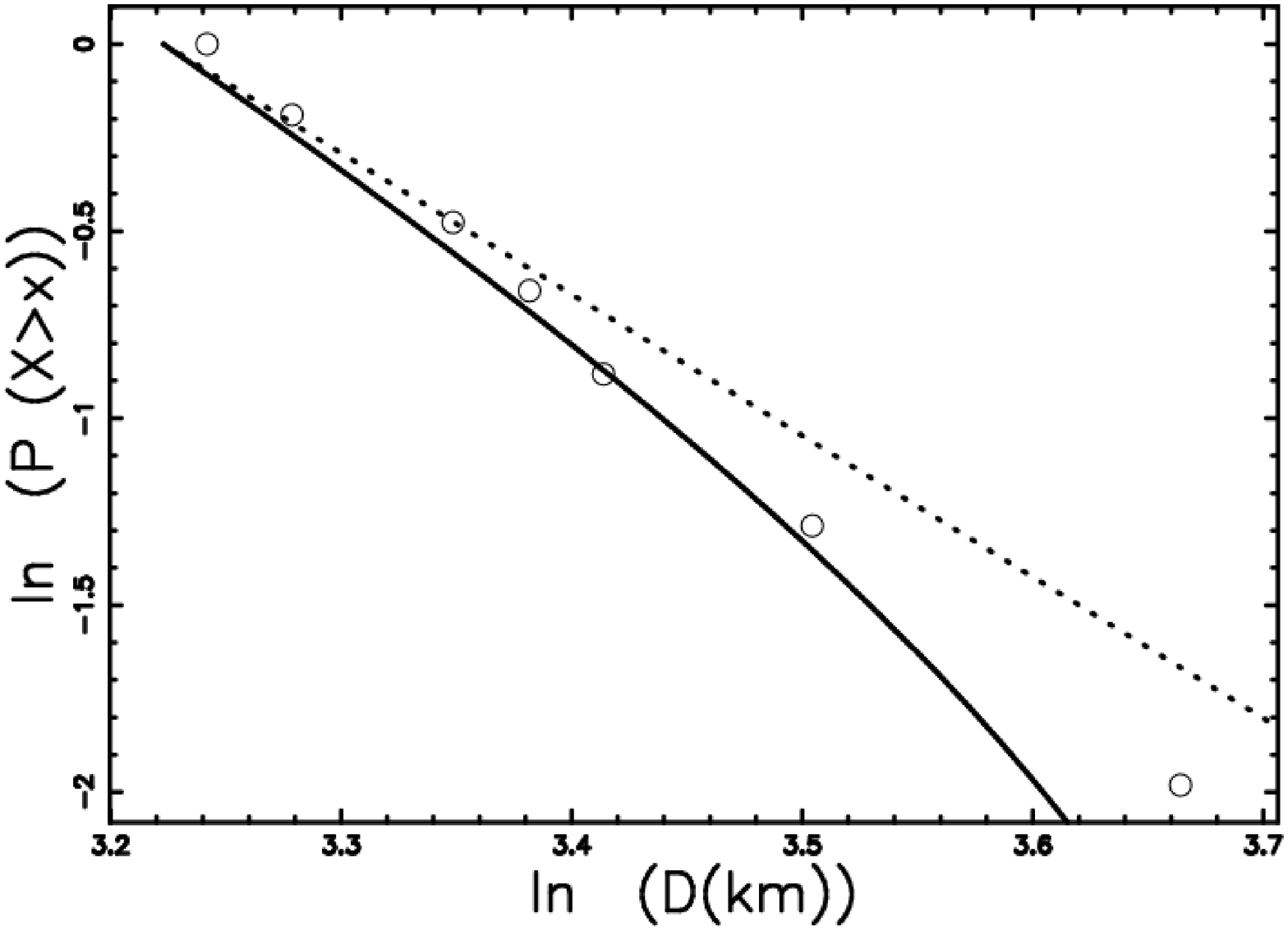}
}
\caption{
ln--ln plot  of the survival function 
of  diameter distribution of the Koronis  Family:
data (empty circles),
survival function  of the truncated Pareto  pdf (full line) and
survival function  of the           Pareto  pdf (dotted line).
A  complete sample is considered with
parameters as in Table~\ref{Koronis}.
}
\label{pareto_tronc_koronis_diam}
\end{figure}

\begin{figure}
{
\includegraphics[width=10cm]{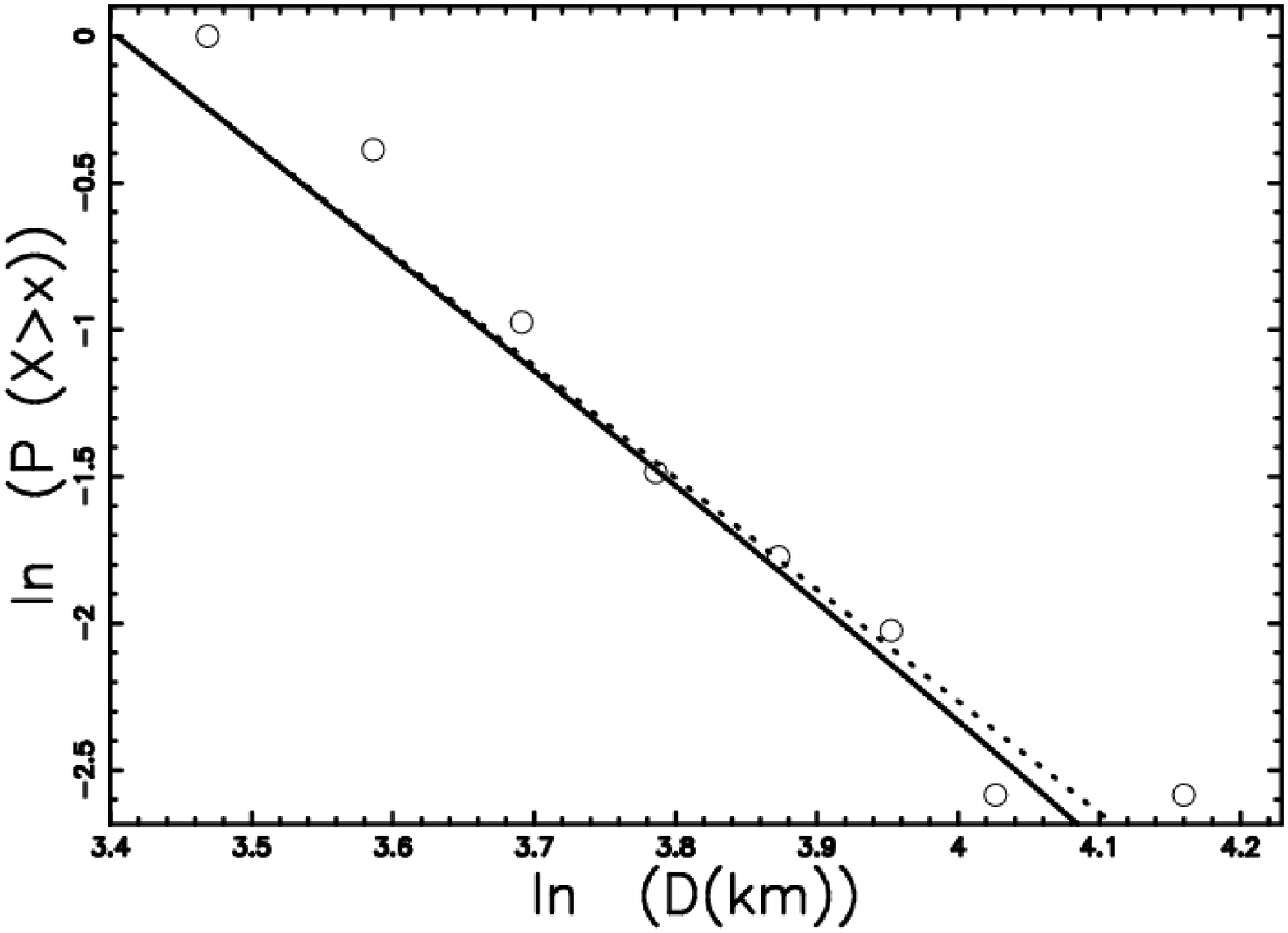}
}
\caption{
ln--ln plot  of the survival function 
of  diameter distribution of the Eos Family:
data (empty circles),
survival function  of the truncated Pareto  pdf (full line) and
survival function  of the           Pareto  pdf (dotted line).
A  complete sample is considered with
parameters as in Table~\ref{Eos}.
}
\label{pareto_tronc_eos_diam}
\end{figure}

\begin{figure}
{
\includegraphics[width=10cm]{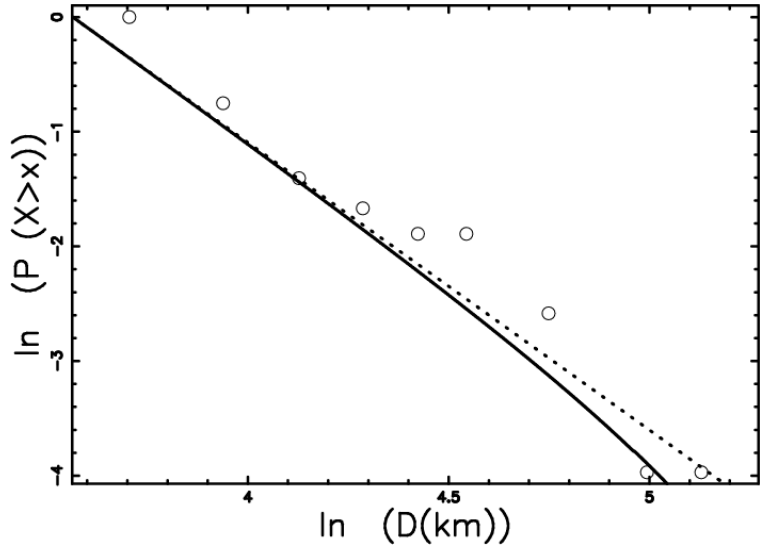}
}
\caption{
ln--ln plot  of the survival function 
of  diameter distribution of the Themis  Family:
data (empty circles),
survival function  of the truncated Pareto  pdf (full line) and
survival function  of the           Pareto  pdf (dotted line).
A  complete sample is considered with
parameters as in Table~\ref{Themis}.
}
\label{pareto_tronc_themis_diam}
\end{figure}

The  sample parameter of the families are 
reported in Table~\ref{Koronis},Table~\ref{Eos} and Table~\ref{Themis} ,
 whereas 
Figure~\ref{pareto_tronc_koronis_diam},
Figure~\ref{pareto_tronc_eos_diam},
Figure~\ref{pareto_tronc_themis_diam} 
report the graphical display of data and the fitting distributions.

\begin{table}[h]
\caption{\it Coefficients of  diameter  distribution of the Koronis family .
         The parameter $c$ is derived through MLE and p=0.033 .}
\label{Koronis}
\begin{tabular}{ccccc}
\hline
a [km]         &  b [km]      & c       &  n  &  $P(X>x)$                \\
\hline
25.1           &  44.3        & 3.77    &  29 & truncated ~Pareto                    \\
\hline
25.1           & $\infty$     & 5.04    &  29 & Pareto                    \\
\hline
\hline
\hline
\end{tabular}
\end{table}

\begin{table}[h]
\caption{\it Coefficients of  diameter  distribution of the Eos family .
         The parameter $c$ is derived through MLE and p=0.681 .}
\label{Eos}
\begin{tabular}{ccccc}
\hline
a [km]         &  b [km]      & c       &  n  &  $P(X>x)$                \\
\hline
30.1           &  110        & 3.80    &  53 & truncated ~Pareto                    \\
\hline
30.1           & $\infty$    & 3.94    &  53 & Pareto                    \\
\hline
\hline
\hline
\end{tabular}
\end{table}

\begin{table}[h]
\caption{\it Coefficients of  diameter  distribution of the Themis family .
         The parameter $c$ is derived through MLE and p=0.67 .}
\label{Themis}
\begin{tabular}{ccccc}
\hline
a [km]         &  b [km]      & c       &  n  &  $P(X>x)$                \\
\hline
35.3           &  249         & 2.5     &  53  & truncated ~Pareto                    \\
\hline
35.3           & $\infty$     & 2.6     &  53 & Pareto                    \\
\hline
\hline
\hline
\end{tabular}
\end{table}

In case of the Koronis family $P_T$ fits the data better than 
$P$ and indeed $p=0.039$ is correspondingly small, whereas 
for the Eos family,
$P$ performs slightly better than $P_T$ (p=0.68), and the estimated 
of $c$ are very closed in both cases. Finally in the third case, the Themis 
family, the two distributions are the same, due to the fact that 
the ratio $a/b=0.14$ is small.

\begin{figure}
{
\includegraphics[width=10cm]{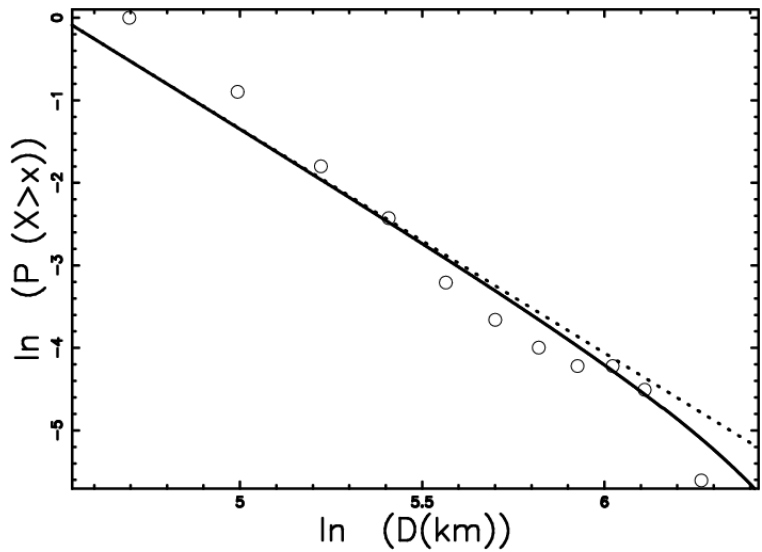}
}
\caption{
ln--ln plot  of the survival function 
of  diameter distribution 
of the  asteroids
extracted from   Astorb database with diameter $> 90$  $Km$,
data (empty circles),
survival function  of the truncated Pareto  pdf (full line) and
survival function  of the           Pareto  pdf (dotted line).
A  complete sample is considered with
parameters as in Table~\ref{all}.
}
\label{pareto_tronc_asteroidimaggioreuno}
\end{figure}

Another interesting catalog is the Asteroid Orbital Elements Database
(Astorb)  which is  visible  at  http://vizier.u-strasbg.fr/;
the  sample parameter of the asteroids  with diameter $> 90$  $Km$
is reported in Table~\ref{all}
and the  fitting survival  function in 
Figure~\ref{pareto_tronc_asteroidimaggioreuno}.

\begin{table}[h]
\caption{\it 
         Coefficients of  diameter  distribution of the  asteroids
         extracted from   Astorb database with diameter $> 90$  $Km$. 
         The parameter $c$ is derived through MLE and p=0.53 .}
\label{all}
\begin{tabular}{ccccc}
\hline
a [km]         &  b [km]      & c       &  n  &  $P(X>x)$                \\
\hline
90.59           &  848.4       & 2.71   &  272 & truncated ~Pareto                    \\
\hline
90.59           & $\infty$     & 2.75    & 272 & Pareto                    \\
\hline
\hline
\hline
\end{tabular}
\end{table}

\section{Simulating  Pareto  tails}
\label{tails}
Two models   that  explains  the Pareto tails 
are  presented.
The first analyzes  the  possibility that the asteroids 
are formed from smaller bodies  and the second one 
introduces a fragmentation model  at the 
light    of the Voronoi diagrams.

\subsection{Accretion}

As a  simple example of how 
a distribution with power can be generated, consider   
the growth of a primeval nebula via accretion, 
that is the process by which nebulae 
``capture'' mass. 
 We start by considering  an uniform pdf  for the initial mass
of $N$ primeval nebulae, $m$, in a  range 
  $m_{min} <    m \leq  m_{max} $ .
At each interaction the $i$-th nebula has a probability $\lambda_i$ 
to increase its mass $m_i$
 that is given by
\begin{equation}
\label{eq:trans}
\lambda_i=(1-\exp(-ak m_i)),
\end{equation}
where $ak$ is a parameter of the simulation;
thus more ``massive'' nebulae are more likely 
to grow, via accretion. 
The quantity of which the primeval nebula 
can grow varies with time, to take into account that the total mass 
available is limited,
\begin{equation}
\delta m(t)=\delta m(0)\exp(-t/\tau)
\label{eq:dm}
\quad ,
\end{equation}
where $\delta m(0)$ represents the maximum 
mass of exchange and $\tau$ the scaling time of the phenomena.
The simulation proceeds as follows: 
a number $r$,  is randomly chosen in the interval $[0,1]$ 
for each nebula, and, 
if $r < \lambda_i$, the mass $m_i$ is increased by 
$\delta m(t)$, where $t$ denotes the iteration of the process.
The process proceed in parallel : at each 
temporal iteration all the primeval nebulae 
are considered.

\begin{figure}
{
\includegraphics[width=10cm]{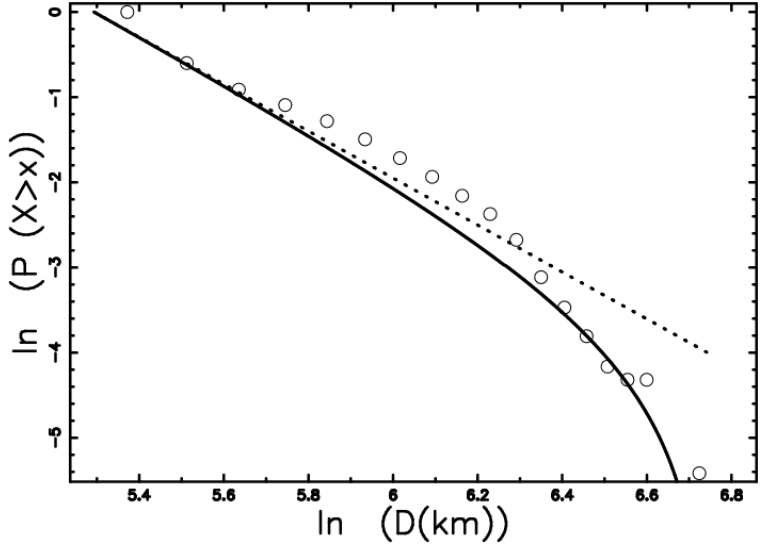}
}
\caption{
log--log plot  of the survival function 
of the   diameter distribution for the primeval nebula.
The truncated Pareto parameters are  $c$=2.75  and p=0.00026~.
}
\label{diameter_asteroids}
\end{figure}

Results of the simulations  have been fitted with both
  Pareto survival distributions.
see Figure~\ref{diameter_asteroids}

\subsection{Fragmentation}

The distribution of fragments size arising from breaking  of  material is
still subject of research as it depends on the actual fragmentation process.
The first model  was developed in \cite{Mott1947}: 
it assumes a time dependent probability of fracture of a
ring after a critical strain has been reached in the material.
The resulting distribution  show that the frequency of occurrence of fragments masses 
follows a cumulative 
distribution of the form :
\begin{equation}
 {N_{m}\over N} =  e^{\displaystyle - {\sqrt {m\over \mu }}} 
\quad ,  
\end{equation}

where $N_m$ is the number of fragments each of whose mass is greater 
or equal to $m$ , $N$ is the total number of fragments ,
\begin{equation}
 \mu = {\overline {m}\over 2 }     \quad ,             
\end{equation}
and $ \overline {m } $ is the averaged mass of a fragment.

\subsubsection {Fractal distribution of fragments size }

In order to generate cells resulting in a fractal distribution of their volumes
 the 
following method can  be adopted, which generalizes the procedure presented in 
\cite{Turcotte1993}.

Consider a domain $\mathcal D$ subdivided into ${N}_0$ 
cubic cells with a linear dimension $l$ that, that in the following, 
 will be referred as zero-order cells.
A zero-order  cell    
is divided into  $k^3$ smaller cubes called zero-order elements, 
with linear dimension $l/k$ and  volumes  
\begin{equation}
V_1=\frac{V_0}{k^3},
\label{volum}
\end{equation}

where $V_0$ is the volume of zero-order cell.
If$P$ is the probability of a zero-order cell to  be fragmented, 
the number $N_1$ of zero-order elements generated  by 
fragmentation is given by 
\begin{equation}
N_1= P \cdot k^3N_0,
\label{new}
\end{equation}

and the number $N_{0a}$
of zero-order cells that have been not fragmented is
\begin{equation}
N_{0a}=(1-P)N_0.
\label{nofragm}
\end{equation}

Each zero-order element now becomes a first-order cell
that can be fragmented into first-order elements
of volumes 
\begin{equation}
V_2=\frac{V_0}{(k^3)^2}
\label{volum2}
\end{equation}
and the number of fragmented  first-order elements
is 
\begin{equation}
N_2=Pk^3N_1=(k^3P)^2N_0.
\label{secord}
 \end{equation}

The number $N_{1a}$ of first order cells that
 have not been fragmented is given by 

\begin{equation}
N_{1a}=k^3P(1-P)N_0.
\label{nofragm1}
\end{equation}

Now first-order elements can be considered second-order cells  
and the procedure repeats itself. The 
volume of the $n$th-order cell $V_n$ is 
\begin{equation}
V_n =  \frac{ V_0}{k^{3n}}
\label{volumn}
\quad ,
\end{equation}

and, after fragmentation, 
the number of $n$th-order cells $N_{na}$ is 

\begin{equation}
N_n = (P k^3)^nN_{0a}= (P k^3)^n( 1- P)N_0  
\label{norder}
\end{equation}

Taking the natural logarithm Eqs. (\ref{volumn}) and (\ref{norder})
leads to 
\begin{equation}
\ln {V_n \over V_0} = - n \ln (k^3),
\label{veg}
\end{equation} 
\begin{equation}
\label{eg}
\ln {N_{na} \over N_{0a}} = - n\ln (Pk)^3
\quad .
\end{equation} 

>From Eqs. (\ref{eg}) and  (\ref{eg}) it is straightforward to  obtain, by elimination of $n$ :
\begin{equation}
{N_{na} \over N_{0a}} = \bigl [{V_n \over V_0}] \displaystyle
^{- {\displaystyle  \ln [P k^3] \over \displaystyle \ln [k^3] }},
\end{equation} 

 that is a fractal distribution with dimension $D$ given by 
\begin{equation}
D  = {3 \ln(P k^3) \over { \ln ( k^3) }} 
\quad .
\end{equation}

\begin{figure}
{
\includegraphics[width=10cm]{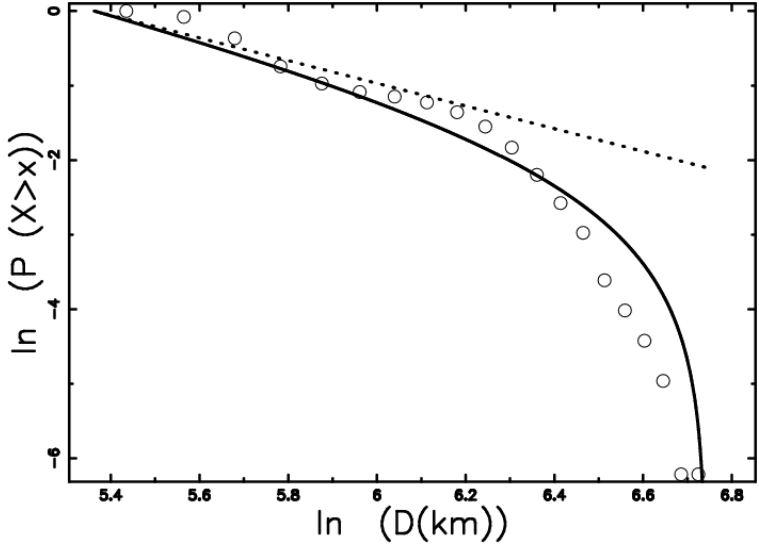}
}
\caption{
log--log plot  of the survival function 
of the   diameter distribution as given by the 
Voronoi Diagrams  in presence of 1000 fractal  seeds.
The  parameters of the simulation 
are  $k$ =2 , $P$ = 0.92  
and therefore $D_{fr}$=2.80.
The truncated Pareto parameters are  $c$=1.52  and 
$p=0 $.}
\label{voronoi_pareto_surv}
\end{figure}

Now we can consider  the center of cells, of any order, 
  as seeds for the generation of
a Voronoi diagram as shown in  
see Figure~ \ref{voronoi_pareto_surv}.

\section{Conclusions}

In this chapter  statistical parameters 
for the truncated  Pareto distribution, namely   
average, variance, median and 
$n-th$ central moment have been calculated.
 Furthermore, also  distribution function and  survival function 
have been derived.
These  quantities  allow  to fit the various  
families of asteroids and the  
Astorb database which  are characterized by   a finite
rather than infinite maximum diameter.
Two possible  simulations  are  suggested to produce  
Pareto tails.
The first one  
results from 
a simple growth process, in which the increase 
of the state variable  
(here mass) depends on the values taken in 
the previous state.
The  second  one is a fragmentation process given
by  3D Voronoi volumes with  a 
fractal  distribution of seeds.


\begin{thebibliography}{10}
\expandafter\ifx\csname url\endcsname\relax
  \def\url#1{\texttt{#1}}\fi
\expandafter\ifx\csname urlprefix\endcsname\relax\def\urlprefix{URL }\fi

\bibitem{Davis1989}
D.~R. {Davis}, S.~J. {Weidenschilling}, P.~{Farinella}, P.~{Paolicchi}, R.~P.
  {Binzel}, {Asteroid collisional history - Effects on sizes and spins}, in:
  {R.~P.~Binzel, T.~Gehrels, \& M.~S.~Matthews} (Ed.), Asteroids II, 1989, pp.
  805--826.

\bibitem{Davis2002b}
D.~R. {Davis}, D.~D. {Durda}, F.~{Marzari}, A.~{Campo Bagatin},
  R.~{Gil-Hutton}, Asteroids III  (2002) 545--558.

\bibitem{Tedesco2005}
E.~F. {Tedesco}, A.~{Cellino}, V.~{Zappal{\'a}}, \aj 129 (2005) 2869--2886.

\bibitem{Michtchenko2010}
T.~A. {Michtchenko}, D.~{Lazzaro}, J.~M. {Carvano}, S.~{Ferraz-Mello}, \mnras
  401 (2010) 2499--2516.

\bibitem{Pareto}
V.~{Pareto}, Cours d' economie politique, Rouge, Lausanne, 1896.

\bibitem{evans}
M.~{Evans}, N.~{Hastings}, B.~{Peacock}, Statistical Distributions - third
  edition, John Wiley \& Sons Inc, New York, 2000.

\bibitem{Cohen1988}
A.~{Cohen}, B.~{Whitten}, Parameter Estimation in reliability and Life Span
  Models, Marcel Dekker, New York, 1988.

\bibitem{Devoto1998}
D.~{Devoto}, S.~{Martinez}, Mathematical Geology 30~(6) (1998) 661 -- 673.

\bibitem{Aban2006}
I.~{Aban}, M.~{Meerschaert}, A.~{Panorska}, Journal of the American Statistical
  Association 101 (2006) 270--277.

\bibitem{Abramowitz1965}
M.~{Abramowitz}, I.~A. {Stegun}, {Handbook of mathematical functions with
  formulas, graphs, and mathematical tables}, Dover, New York, 1965.

\bibitem{Seggern1992}
D.~{von Seggern}, CRC Standard Curves and Surfaces, CRC, New York, 1992.

\bibitem{Thompson1997}
W.~J. {Thompson}, Atlas for computing mathematical functions,
  Wiley-Interscience, New York, 1997.

\bibitem{Grad2000}
I.~{Gradshteyn}, I.~{Ryzhik }, Table of Integrals, Series, and Products,
  Academic Press, San Diego, 2000.

\bibitem{Prud1986}
A.~{Prudnikov}, O.~{Brychkov }, Y.~{Marichev }, Integrals and Series, Gordon
  and Breach Science Publishers, Amsterdam, 1986.

\bibitem{Ali2006}
M.~{Masoom Ali}, S.~{Nadarajah}, Computer Communications 30 (2006) 1--4.

\bibitem{Mott1947}
N.~F. {Mott}, Royal Society of London Proceedings Series A 189 (1947) 300--308.

\bibitem{Turcotte1993}
{{Turcotte}, D. L.}, {Fractals and chaos in geology and geophysics.},
  {Cambridge University Press. }, Cambridge, 1993.

\end{thebibliography}

\end{document}